\documentclass[notitlepage,aps,prd,a4paper,11pt,amsfonts,amssymb,amsmath]{revtex4-1}
\usepackage{graphicx}
\usepackage{bigints}

\begin{document}
\title{Traversable wormholes without exotic matter in multimetric repulsive gravity}

\author{Manuel Hohmann}
\email{manuel.hohmann@ut.ee}
\affiliation{Teoreetilise F\"u\"usika Labor, F\"u\"usika Instituut, Tartu \"Ulikool, Riia 142, 51014 Tartu, Estonia}

\begin{abstract}
We present a static, spherically symmetric, traversable wormhole solution to multimetric gravity which is sustained by only non-exotic matter, i.e., matter which satisfies all energy conditions. The possibility of this solution arises from the fact that under certain conditions the multimetric gravitational field equations reduce to the Einstein equations, but with a negative effective gravitational constant. We show that the Arnowitt-Deser-Misner mass of this wormhole vanishes, so that it appears massless to observers in the asymptotically flat spacetime. We finally speculate on the feasibility of creating and maintaining this type of wormhole by an advanced civilization.
\end{abstract}
\maketitle

\section{Introduction}
It is a well-known fact that solutions to general relativity which describe traversable wormholes require the existence of exotic matter, i.e., matter which violates the null energy condition (NEC)~\cite{Visser:1995cc}. The latter states that for any null vector \(k^a\), the energy-momentum tensor \(T_{ab}\) satisfies \(T_{ab}k^ak^b \geq 0\). The violation of the NEC is a consequence of the ``flaring-out condition''~\cite{Visser:1995cc} which follows from the existence of a wormhole throat where its radius is minimal, and where \(G_{ab}k^ak^b < 0\) for some null vector \(k^a\). The Einstein equations \(G_{ab} = 8\pi GT_{ab}\) then imply that also \(T_{ab}k^ak^b < 0\). This was first shown for static, spherically symmetric wormholes~\cite{Morris:1988cz,Morris:1988tu} and later generalized to non-static, non-symmetric wormholes~\cite{Hochberg:1997wp,Hochberg:1998ii,Hochberg:1998ha}. However, this result crucially depends on the gravitational field equations, and indeed in modified gravity theories the requirement of exotic matter can be minimized or even completely avoided~\cite{Lobo:2007qi,Lobo:2008zu,Lobo:2009ip,Garcia:2010xb,MontelongoGarcia:2010xd,Harko:2013yb}. In these theories the NEC, being a condition on the matter energy-momentum tensor, does not translate to a geometric constraint in the same way as in general relativity, and instead leads to possibly weaker conditions on the allowed geometries~\cite{Capozziello:2013vna}. The aim of this article is to show that wormhole solutions without exotic matter also exist in multimetric gravity theories. In the following we present a simple, static, spherically symmetric, traversable wormhole solution and show that it satisfies all energy conditions.

\section{Multimetric gravity}
In this article we study wormholes in the context of a recently discussed class of multimetric gravity theories containing \(N \geq 2\) metric tensors \(g^I_{ab}, I = 1, \ldots, N\) and a corresponding number of standard model copies \(\varphi^I\)~\cite{Hohmann:2013oca}. The action of these theories is of the common form
\begin{equation}\label{eqn:actionsplit}
S = S_G[g^1, \ldots, g^N] + \sum_{I = 1}^{N}S_M[g^I, \varphi^I]\,,
\end{equation}
where \(S_M[g^I, \varphi^I]\) denotes the standard model action and the gravitational part of the action is given by
\begin{equation}\label{eqn:repaction}
S_G = \frac{1}{16\pi}\bigintss\left(c_1\sum_{I = 1}^{N}R^I + c_2\sum_{I,J = 1}^{N}g^{I\,ab}R^J{}_{ab} + Q(S)\right)\omega\,.
\end{equation}
Here \(c_{1,2}\) are constants, \(\omega\) denotes the mixed volume form
\begin{equation}
\omega = d^4x\prod_{I = 1}^{N}\left(g^I\right)^{\frac{1}{2N}}
\end{equation}
and \(Q(S)\) is quadratic in the connection difference tensors \(S^{IJ} = \Gamma^I - \Gamma^J\) formed from the Christoffel symbols of the different metrics. It follows from the structure of the action~\eqref{eqn:actionsplit} that each standard model copy couples only to its own metric tensor \(g^I_{ab}\), and therefore obeys a well-defined causality. It moreover follows that the interaction between the different standard model copies is mediated only through gravity, so that they mutually appear dark. This can also be seen from the gravitational field equations, which are derived by variation of the total action with respect to the metric tensors. They take the form
\begin{equation}\label{eqn:multieom}
K^I_{ab}[g^1, \ldots, g^N] = 8\pi T^I_{ab}[g^I, \varphi^I]\,,
\end{equation}
where \(T^I_{ab}\) are the usual standard model energy-momentum tensors derived from the matter action \(S_M\) and the curvature tensors \(K^I_{ab}\) are analogously derived from the gravitational action \(S_G\). A closer look at the terms in the gravitational part~\eqref{eqn:repaction} of the action further shows that \(K^I_{ab}\) is of second derivative order and contains only terms which are built from the Ricci tensors \(R^I_{ab}\) and originate from the first two terms in \(S_G\), as well as terms which are either quadratic in the connection difference tensors \(S^{IJ}\) or linear in their covariant derivatives \(\nabla^IS^{JK}\) with respect to the Levi-Civita connection \(\nabla^I\) of \(g^I_{ab}\).

We further restrict ourselves to theories which employ the Copernican principle in the sense that the action and in consequence also the field equations are symmetric with respect to permutations of the \(N\) sectors. In the Newtonian limit the gravitational field equations then yield the most general Poisson equations compatible with this symmetry,
\begin{equation}\label{eqn:poisson}
-\frac{1}{4\pi}\nabla^2\Phi^I = G\rho^I + G^*\sum_{J \neq I}\rho^J\,,
\end{equation}
where \(\rho^I\) are the matter densities and the Newtonian potentials \(\Phi^I\) enter the metric in the form \(g^I_{00} = -1 + 2\Phi^I\). The constants appearing here are the usual Newtonian constant \(G\), which measures the gravitational force within each of the sectors and which we rescale to \(G = 1\) by a choice of units, and a second Newtonian constant \(G^*\) measuring the gravitational force between different standard model copies. This Newtonian limit is implemented if the constants \(c_{1,2}\) satisfy
\begin{equation}\label{eqn:constants}
c_1 + c_2 = \frac{1}{1 + (N - 1)G^*}\,.
\end{equation}
In order to elucidate this we explicitly write out the simplest possible case \(N = 2\), where the Poisson equation~\eqref{eqn:poisson} takes the form
\begin{equation}\label{eqn:poisson2}
-\frac{1}{4\pi}\nabla^2\left(\begin{array}{c}
\Phi^1\\
\Phi^2
\end{array}\right) = \left(\begin{array}{cc}
1 & G^*\\
G^* & 1
\end{array}\right)\left(\begin{array}{c}
\rho^1\\
\rho^2
\end{array}\right)
\end{equation}
in units where \(G = 1\), and is achieved if the constants~\eqref{eqn:constants} satisfy \(c_1 + c_2 = (1 + G^*)^{-1}\). Note that this relation excludes the value \(G^* = -1\), for which the two matter types \(\rho^1, \rho^2\) would repel each other with equal strength compared to the gravitational attraction within each of the two sectors. This exclusion has been proven earlier in terms of a no-go theorem~\cite{Hohmann:2009bi}. The theorem states at its core that the matrix on the right hand side of the Poisson equation~\eqref{eqn:poisson2} must be invertible, which is equivalent to the right hand side of~\eqref{eqn:constants} being non-singular. This no-go theorem together with the intent to examine also the simple choice \(G^* = -1\) are the reason why we extend our discussion to the more general multimetric case \(N \geq 2\) here.

We finally remark that experimental consistency at the post-Newtonian level further requires a non-vanishing term \(Q(S)\) and restricts its form~\cite{Hohmann:2013oca}. However, we will not discuss these constraints here since they will turn out to be irrelevant for the remainder of this article.

\section{Symmetric solutions}
We now turn our focus to a particular class of solutions to multimetric gravity which are symmetric under permutations of the sectors, i.e., in which all metrics \(g^I_{ab}\) are equal. It then follows from our assumption on the symmetry of the field equations~\eqref{eqn:multieom} that also all energy-momentum tensors \(T^I_{ab}\) are equal. We further see from the structure of the gravitational action~\eqref{eqn:repaction} that the field equations simplify in the case of equal metrics: all connection difference tensors \(S^{IJ}\) vanish, all Ricci tensors \(R^I_{ab}\) are equal and the remaining terms take the simple form
\begin{equation}\label{eqn:symeom}
G_{ab} = 8\pi G_{\text{eff}}T_{ab}\,.
\end{equation}
Here we have already dropped the sector indices \(I\), since also all Einstein tensors \(G^I_{ab}\) are equal, and introduced an effective gravitational constant \(G_{\text{eff}} = (c_1 + c_2)^{-1}\). We have thus reproduced the Einstein equations, up to an additional factor. Hence, every solution to general relativity induces a symmetric solution to our multimetric gravity theories, where \(T_{ab}\) is to be replaced by \(G_{\text{eff}}T_{ab}\).

The effective gravitational constant appearing here deserves particular attention. It follows from the relation~\eqref{eqn:constants} that \(G_{\text{eff}}\) becomes negative provided that the gravitational interaction between different standard model copies is repulsive and of sufficient strength,
\begin{equation}\label{eqn:repgrav}
G^* < \frac{1}{1 - N}\,.
\end{equation}
In the simple bimetric case \(N = 2\) this condition is satisfied for \(G^* < -1\), which means that the mutual repulsion between different standard model copies must be stronger than the gravitational attraction within each matter sector. Another arguably simple case is to fix \(G^* = -1\), which then requires the number of metrics to be \(N > 2\). The latter choice has been considered earlier in the context of cosmology, where the negative effective gravitational constant in the simplified field equations~\eqref{eqn:symeom} yields an accelerating expansion of the universe~\cite{Hohmann:2010vt}. We remark that this accelerating expansion can also be achieved in the bimetric case \(N = 2\) with \(G^* < -1\) since it depends only on the sign of the effective gravitational constant \(G_{\text{eff}}\).

In the remainder of this article we will consider both cases, since it will turn out that \(G_{\text{eff}}\) is the only crucial ingredient for our discussion of the energy conditions, so that any choice of \(N \geq 2\) is allowed.

\section{Wormhole metric}
\begin{figure}
\centering
\includegraphics[width=0.5\textwidth]{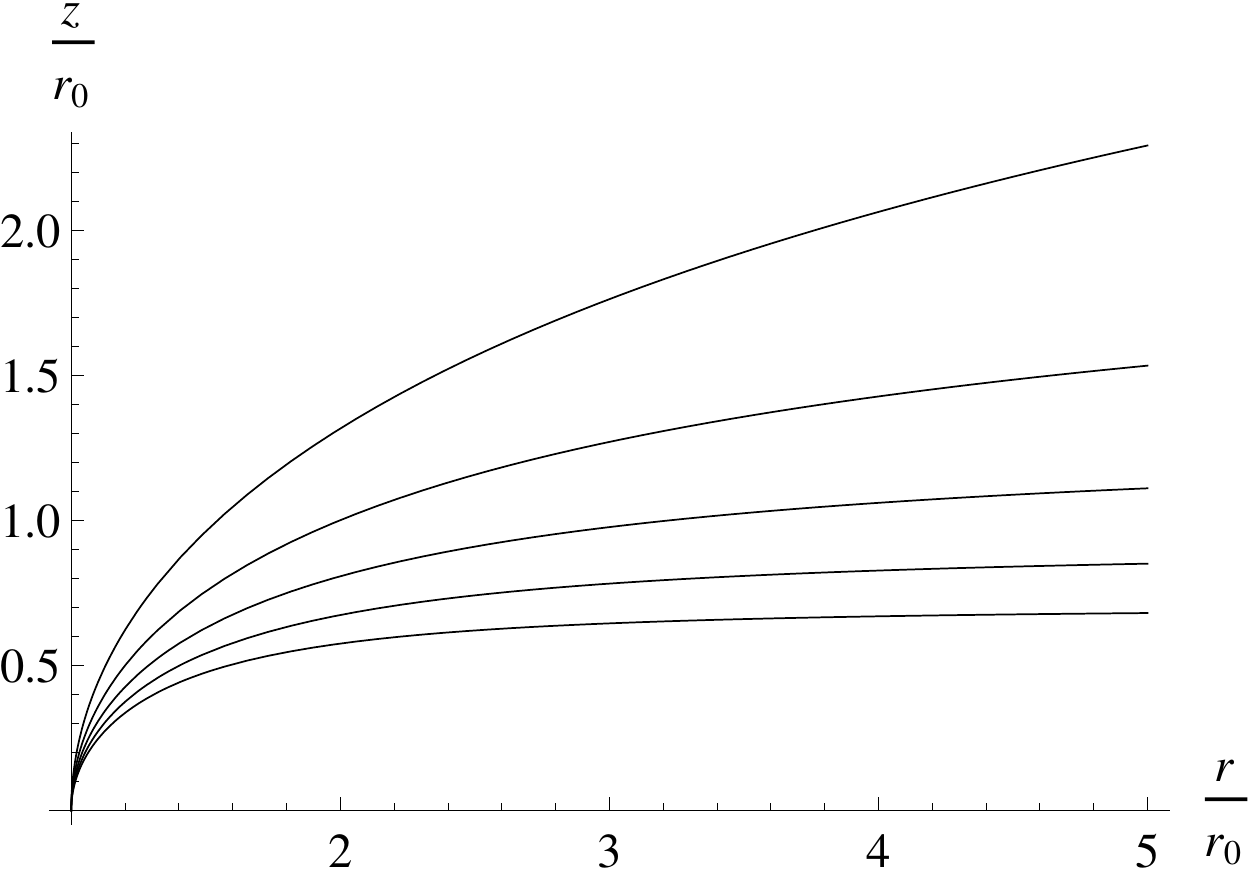}
\caption{Embedding of the wormhole in $4+1$-dimensional flat spacetime for \(n = 1, \ldots, 5\) from top to bottom. The plot shows only the coordinates \(r\) and \(z\) for the upper end \(z \geq 0\) of the wormhole.}\label{fig:whembed}
\end{figure}

The solution we discuss in the following is the static, spherically symmetric, traversable wormhole connecting two asymptotically Minkowski spacetimes, which is given by the metric~\cite{Morris:1988cz}
\begin{equation}\label{eqn:whmetric}
g_{ab}dx^adx^b = -e^{2\Phi(r)}dt^2 + \frac{dr^2}{1 - b(r) / r} + r^2(d\theta^2 + \sin^2\theta\,d\phi^2)\,.
\end{equation}
The radial coordinate \(r\) we use here is chosen so that the circumference of a circle around the center of the wormhole is given by \(2\pi r\). The spacetime manifold consists of two charts in both of which \(r \geq r_0\) for a constant \(r_0 > 0\), and which are glued together at the wormhole throat \(r = r_0\). For the functions \(\Phi(r)\) and \(b(r)\), called the redshift and shape functions, we choose
\begin{equation}\label{eqn:multiwh}
\Phi(r) = 0\,, \quad b(r) = r_0^{n + 1}/r^n
\end{equation}
with \(n \geq 1\). One easily checks that the spacetime defined by this choice is asymptotically flat, since \(b/r \to 0\) and \(\Phi \to 0\) as \(r \to \infty\). It does not possess any horizons, since \(\Phi\) is everywhere finite. Finally, from the flaring-out condition \(b - b'r > 0\) follows that it has indeed the shape of a wormhole~\cite{Morris:1988cz}. The latter can also be seen from an isometric embedding of the wormhole spacetime in a flat $4+1$-dimensional spacetime. Denoting the additional spatial coordinate by \(z\), the embedded hypersurface is given by
\begin{equation}
z(r) = \pm\int_{r_0}^{r}\frac{dr'}{\sqrt{r'/b(r') - 1}} = \mp r_0\left[\frac{1}{n + 1}\mathrm{B}\left(\left(\frac{r_0}{r}\right)^{n + 1}; \frac{1}{2} - \frac{1}{n + 1}, \frac{1}{2}\right) - \sqrt{\pi}\frac{\Gamma\left(\frac{1}{2} - \frac{1}{n + 1}\right)}{\Gamma\left(-\frac{1}{n + 1}\right)}\right]\,,
\end{equation}
where \(\Gamma\) denotes the Euler gamma function and \(\mathrm{B}\) denotes the incomplete Euler beta function
\begin{equation}
\mathrm{B}(x; a, b) = \int_{0}^{x}t^{a - 1}(1 - t)^{b - 1}dt\,.
\end{equation}
The upper half \(z \geq 0\) of this embedding is shown in figure~\ref{fig:whembed}.

\section{Energy conditions}
Inserting the metric~\eqref{eqn:whmetric} into the symmetric gravitational field equations~\eqref{eqn:symeom} we find the energy-momentum tensor
\begin{equation}\label{eqn:whemt}
T_{ab}dx^adx^b = \frac{1}{8\pi G_{\text{eff}}}\left(\frac{r_0}{r}\right)^{n + 1}\left(-\frac{n}{r^2}dt^2 + \frac{dr^2}{r_0^{n + 1}/r^{n - 1} - r^2} + \frac{n + 1}{2}(d\theta^2 + \sin^2\theta\,d\phi^2)\right)\,.
\end{equation}
In order to test whether the energy conditions are satisfied we use the tangent space basis
\begin{equation}
e_{\hat{t}} = e^{-\Phi}\partial_t\,, \quad e_{\hat{r}} = \sqrt{1 - b/r}\,\partial_r\,, \quad e_{\hat{\theta}} = r^{-1}\partial_{\theta}\,, \quad e_{\hat{\phi}} = (r \sin\theta)^{-1}\partial_{\phi}\,,
\end{equation}
which is orthonormal with respect to the metric~\eqref{eqn:whmetric}. In this basis the non-vanishing components of the energy-momentum tensor~\eqref{eqn:whemt},
\begin{equation}\label{eqn:whevs}
T_{\hat{t}\hat{t}} = \rho = -\frac{n}{8\pi G_{\text{eff}}}\frac{r_0^{n + 1}}{r^{n + 3}}\,, \quad T_{\hat{r}\hat{r}} = p_r = \frac{\rho}{n}\,, \quad T_{\hat{\theta}\hat{\theta}} = T_{\hat{\phi}\hat{\phi}} = p_l = -\frac{n + 1}{2n}\rho\,,
\end{equation}
are the matter density \(\rho\), the radial pressure \(p_r\) and the lateral pressure \(p_l\). Using these variables we find the expressions
\begin{equation}
\rho + p_r \geq 0\,, \quad \rho + p_l \geq 0
\end{equation}
for the null energy condition,
\begin{equation}
\rho \geq 0\,, \quad \rho + p_r \geq 0\,, \quad \rho + p_l \geq 0
\end{equation}
for the weak energy condition,
\begin{equation}
\rho \geq 0\,, \quad -\rho \leq p_r \leq \rho\,, \quad -\rho \leq p_l \leq \rho\,,
\end{equation}
for the dominant energy condition and
\begin{equation}
\rho + p_r + 2p_l \geq 0\,, \quad \rho + p_r \geq 0\,, \quad \rho + p_l \geq 0\,.
\end{equation}
for the strong energy condition~\cite{Wald:1984rg}. One can immediately read off that all energy conditions are satisfied for the values~\eqref{eqn:whevs} provided that the matter density \(\rho\) is positive. This is the case for a negative effective gravitational constant \(G_{\text{eff}}\). The negative factor in the symmetric gravitational field equations~\eqref{eqn:symeom} therefore allows us to sustain a traversable wormhole using only ordinary, non-exotic matter.

\section{ADM mass}
As another interesting property we calculate the Arnowitt-Deser-Misner (ADM) mass~\cite{Arnowitt:1962hi} of our wormhole solution. For the metric~\eqref{eqn:whmetric} we find
\begin{equation}
M = \lim_{r \to \infty}\frac{b/2}{1 - b/r} = 0\,,
\end{equation}
where the last equality follows from the definition~\eqref{eqn:multiwh} of the shape function \(b\). The (undisturbed) wormhole therefore appears massless to observers in the two asymptotically Minkowski spacetimes. This is no longer true if we consider massive objects passing through the wormhole~\cite{Visser:1995cc}; however, this goes beyond the scope of this article.

\section{Construction and stability}
Given a wormhole solution which is sustained by only non-exotic matter, one may speculate on the possibility to create and maintain such a wormhole. Unfortunately already a simple thought experiment shows that the construction of a multimetric wormhole is beyond any practical feasibility, and unlikely to happen naturally. The main problem arises from the fact that in order to work with the simplified field equations~\eqref{eqn:symeom} we needed to construct a solution in which all metrics \(g^I_{ab}\), and therefore the energy-momentum tensors \(T^I_{ab}\) for all standard model copies, are equal. We further needed to assume that different standard model copies repel each other, so that the effective gravitational constant \(G_{\text{eff}}\) in the field equations~\eqref{eqn:symeom} becomes negative. It thus follows that the assembly of a wormhole sustained by equal amounts of all \(N\) matter types is obstructed by their mutual repulsion, and thus unlikely to occur naturally. Also for a civilization constituted by one of the standard model copies the construction of a multimetric wormhole is practically impossible, since the only way to interact with and assemble the other \(N - 1\) matter types, which appear dark to this civilization, is by means of gravity. It would rather need the cooperation of \(N\) civilizations, each constituted by a different matter type, coordinating their efforts using gravitational communication, to effectively manipulate the sustaining matter of the wormhole. Finally, even if these civilizations should manage to assemble a wormhole, it is very likely to be unstable against fluctuations, since any shift of the different matter types with respect to each other would cause them to further separate and disrupt the wormhole due to their mutual repulsion.

We conclude that even though non-exotic wormhole solutions to multimetric gravity theories exist, their construction and maintenance is far from being feasible even by the joint effort of several advanced civilizations. Similar considerations may apply to other spacetime geometries which require exotic matter in general relativity, such as warp drives: even if one finds a warp drive spacetime which does not require exotic matter in multimetric gravity, it will probably suffer from the same drawbacks which obstruct its practical realization.

\acknowledgments
The author is happy to thank Laur J\"arv for helpful discussions and comments. He gratefully acknowledges full financial support from the Estonian Research Council through the Postdoctoral Research Grant ERMOS115.

\end{document}